\documentclass[%
 reprint,
 amsmath,amssymb,
 aps,
prb,
superscriptaddress,
longbibliography
]{revtex4-1}

\usepackage{graphicx}
\usepackage{dcolumn}
\usepackage{bm}
\usepackage{bbold}
\usepackage{braket}
\usepackage[ruled,vlined]{algorithm2e} 

\usepackage{hyperref}
\hypersetup{colorlinks, linkcolor={blue},citecolor={blue}, urlcolor={blue}}

\begin{document}

\preprint{APS/123-QED}

\title{Backscatter Communication Through Disordered Media Enabled by a Programmable Leaky Cavity}

\author{Cl{\'e}ment Ferise$^*$}
\affiliation{Univ Rennes, CNRS, IETR - UMR 6164, F-35000 Rennes, France}%
\affiliation{Laboratory of Wave Engineering, {\'E}cole Polytechnique F{\'e}d{\'e}rale de Lausanne (EPFL), 1015 Lausanne, Switzerland}%

\author{Pierre Granier}
\affiliation{Univ Rennes, CNRS, IETR - UMR 6164, F-35000 Rennes, France}%

\author{Antton Goïcoechea}
\affiliation{Univ Rennes, CNRS, IETR - UMR 6164, F-35000 Rennes, France}%

\author{François Sarrazin}
\affiliation{Univ Rennes, CNRS, IETR - UMR 6164, F-35000 Rennes, France}%

\author{Philippe Besnier}
\affiliation{Univ Rennes, CNRS, IETR - UMR 6164, F-35000 Rennes, France}%

\author{Matthieu Davy$^*$}
\affiliation{Univ Rennes, CNRS, IETR - UMR 6164, F-35000 Rennes, France}%
\email{matthieu.davy@univ-rennes.fr}

\date{\today}

\begin{abstract} 
Retrieving information through disordered scattering systems remains a major challenge in wireless communication and sensing due to the severe distortion and attenuation caused by multiple scattering. Here, we demonstrate robust focusing through random media using a leaky cavity made programmable by reconfigurable metasurfaces. We show that the leaky cavity leverages backscattering from the disordered medium to enhance the degree of control over the transmitted field. We achieve an enhancement in focused intensity by an order of magnitude compared to a horn antenna. We further demonstrate robust communication through the medium using low-power amplitude-modulated signals, and finally implement a backscatter communication scheme in which the same cavity illuminates and decodes data from a passive reflectivity-modulating target embedded within the disorder. Our results open new possibilities for reliable communication and sensing in complex environments. 
\end{abstract}

\maketitle

Unlike conventional wireless communication systems, backscatter communications utilize the modulation of the scattering environment rather than the transmitted signal itself to encode information \cite{zhang2017freerider,van2018ambient,niu2019overview,zhao2020metasurface,liang2022backscatter}. By temporarily tuning the impedance of an antenna, the scattering environment can be modulated so that a field carries a message decodable by receiving antennas. This field may originate from non-cooperative sources of illumination \cite{van2018ambient}, such as Wi-Fi signals in indoor settings \cite{zhao2020metasurface}, or from transmitted radio frequency (RF) signals designed to reconstruct the impedance modulation with an enhanced signal-to-noise ratio. Backscattering devices are also particularly valuable for sensing applications, as they can effectively monitor changes in the environment \cite{van2018ambient}.  

While backscatter devices look very promising for applications in the internet of things (IoT) and sensor networks, since they consume significantly less power than transmitting antennas -- no RF signal generation is required -- their applications are strongly limited by the noise level. Interrogating these devices is especially challenging when they are placed behind or within a scattering medium. Wavefronts are typically scrambled by scattering within the environment, yielding a `speckle pattern' : i.e., a random spatial distribution of the field with high and vanishing intensity spots~\cite{rotter2017light}. In this case, the noise fluctuations limit the estimation precision of the target state.

To overcome these limitations, one can leverage wavefront shaping techniques that have garnered significant attention in recent years due to their potential applications in wireless communication, imaging, and energy transfer~\cite{Mosk2012,Vellekoop2008a,vellekoop2010exploiting,rotter2017light,cao2022shaping}. They rely on the precise manipulation of the incident field to control the transmitted or reflected fields to enhance transmission between transmitting and receiving channels~\cite{Gerardin2014,popoff2014coherent}, focus waves on a local target~\cite{derode1995robust,vellekoop2007focusing,vellekoop2010exploiting,horstmeyer2015guidestar}, or deposit energy on an extended area~\cite{cheng2014focusing,jeong2018focusing,bender2022depth},  to cite a few examples. In these applications, the optimal incident wavefield can be determined from the scattering matrix $S(\omega)$, which encodes the complete linear relationship between input and output channels. Wavefront shaping can also be employed to interrogate a backscattering device embedded within a scattering medium. When the device alternates between two states, the optimal incident wavefront for maximizing information retrieval can be derived from a singular value decomposition of the differential scattering matrix $\Delta S$ between the two target states~\cite{bouchet2021optimal,yeo2022time}. In the limit of small perturbations, this optimal wavefront coincides with the one that maximizes the Fisher information, thereby ensuring parameter estimation with ultimate precision~\cite{bouchet2021maximum}.
 

In the microwave regime, wavefront shaping through complex environments can commonly be achieved using transmitting phased arrays, where each antenna element is electronically controlled to produce a desired radiation pattern. An alternative and increasingly attractive approach involves embedding a reconfigurable intelligent surface within a leaky chaotic cavity, excited by a single or a few feed ports \cite{mazloum2024indoor}. By tuning the internal field distribution via the electronically programmable metasurface \cite{PhysRevLett.115.017701,gros2020tuning}, it becomes possible not only to illuminate a scene with diverse random wavefields for applications such as computational imaging \cite{sleasman2020implementation,imani2020review,sleasman2022computational}, but also to synthesize arbitrary complex wavefronts at the cavity’s aperture for advanced wave control—such as focusing energy precisely on a target location \cite{mazloum2024indoor}. This method offers a low-cost and efficient solution that can, in some cases, serve as a substitute for conventional phased arrays. 

Compared to thin dynamic metasurface antennas \cite{yoo2018enhancing,shlezinger2021dynamic},  reconfigurable leaky cavities result in higher quality factors, offering enhanced field confinement and stronger sensitivity to internal reconfiguration. A key feature of this approach indeed lies in the mutual coupling between metasurface elements, which results in a nonlinear mapping between the metasurface pixel states and the output wavefront. This nonlinearity is enhanced by multiple scattering occurring within the chaotic cavity, where waves interact repeatedly with the metasurface. Such multiple interactions significantly enhance the sensitivity of the radiated field to each pixel configuration \cite{prod2024mutual}, thereby increasing the degree of control over the transmitted wavefront. To achieve a target field, iterative optimization procedures can be employed. While these methods do not guarantee convergence to a global optimum, they have proven effective in practice, particularly when initialized with favorable configurations~\cite{vellekoop2008phase,del2016spatiotemporal}.

\begin{figure}
    \centering
    \includegraphics[width=8.5cm]{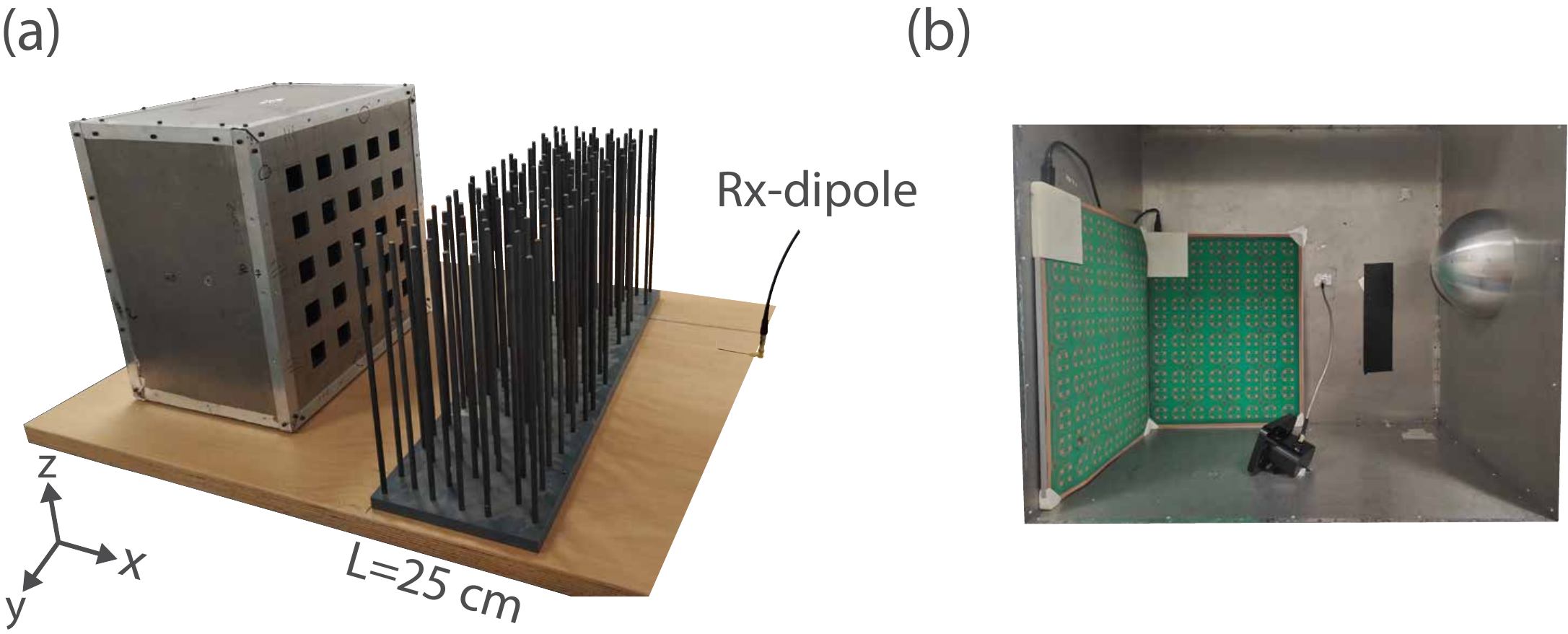}
    \caption{(a) Photograph of the experimental setup for a $L=25$~cm length scattering medium positioned between the transmitting reconfigurable cavity and the receiving dipole antenna. The partially open interface of the cavity consists of a grid of 25 square apertures, each $4~\mathrm{cm}^{2}$. The scattering medium is a collection of aluminum rods. The receiving dipole is located at mid-height of the cavity, i.e., $z\simeq25$~cm, and at a distance $y = 60$~cm from the cavity interface. (b) Photograph of the interior of the cavity showing the two reconfigurable metasurface panels.}
    \label{fig:: Experimental SetUp}
\end{figure}

In this work, we demonstrate improved information retrieval through complex scattering media using a reconfigurable leaky microwave cavity. The cavity, equipped with programmable metasurfaces, enables precise control of the wavefront radiated through its aperture, allowing focused energy delivery to a receiver positioned behind the scattering medium. Remarkably, the presence of the scattering sample enhances the degree of wavefront control: backscattered waves from the medium effectively suppress unstirred components that would otherwise limit focusing performance. We then implement a realistic communication channel, first in a conventional configuration between a transmitter and the reconfigurable cavity, and subsequently in a backscatter communication scenario using an antenna with impedance modulation.

\section{Transmission through a scattering medium}
\label{sec:xp_setup}

The experimental setup consists of a three-dimensional metallic leaky cavity with dimensions of $50 \times 50 \times 30~\mathrm{cm}^{3}$ (see Fig.~\ref{fig:: Experimental SetUp}(a)). The cavity is made chaotic by attaching a metal hemisphere to one of its inner walls. The front interface is partially open with a grid of $25$ apertures {($4~\mathrm{cm}^{2}$ opening)} arranged in a 5-row by 5-column configuration. A single waveguide-to-coax transition is placed inside the cavity and serves as the transmitting source,  operating at a frequency of $\nu_{0}=5.2$~GHz. Additionally, two reconfigurable metasurface panels are positioned inside the cavity to modulate the reflected wavefront. Each metasurface is composed of $152$ one-bit programmable pixels with two distinct electromagnetic responses, phase-shifted by $\pi$. A total number of $304$ pixels can therefore be optimized. A photograph of the metasurfaces placed inside the cavity is shown in Fig~\ref{fig:: Experimental SetUp}(b).

The reconfigurable cavity is positioned with its partially open interface facing a scattering medium. The medium consists of a collection of vertical ($z$-axis) aluminum rods mounted on $25 \times 75$~cm horizontal panels. The overall length, $L$, of the medium is adjustable by varying the number of panels. Three configurations are considered in this work: $L=0$~cm (no scattering panels), $L=25$~cm (one scattering panel), and $L=50$~cm (two scattering panels). The experimental setup is displayed for $L=25$~cm in Fig~\ref{fig:: Experimental SetUp}(a). 
The medium is arranged such that the distance between the cavity interface and the far (rightmost) boundary of the medium remains identical for  $L=25$~cm and $L=50$~cm. Consequently, for $L=25$~cm, the single panel is positioned at the far right, away from the cavity, leaving an empty space between the cavity and the panel; for $L=50$~cm, an additional panel is inserted in front of the cavity to fill this space, while the first one remains fixed at the rightmost position.

\begin{figure*} [t]
    \centering
    \includegraphics[width=18cm]{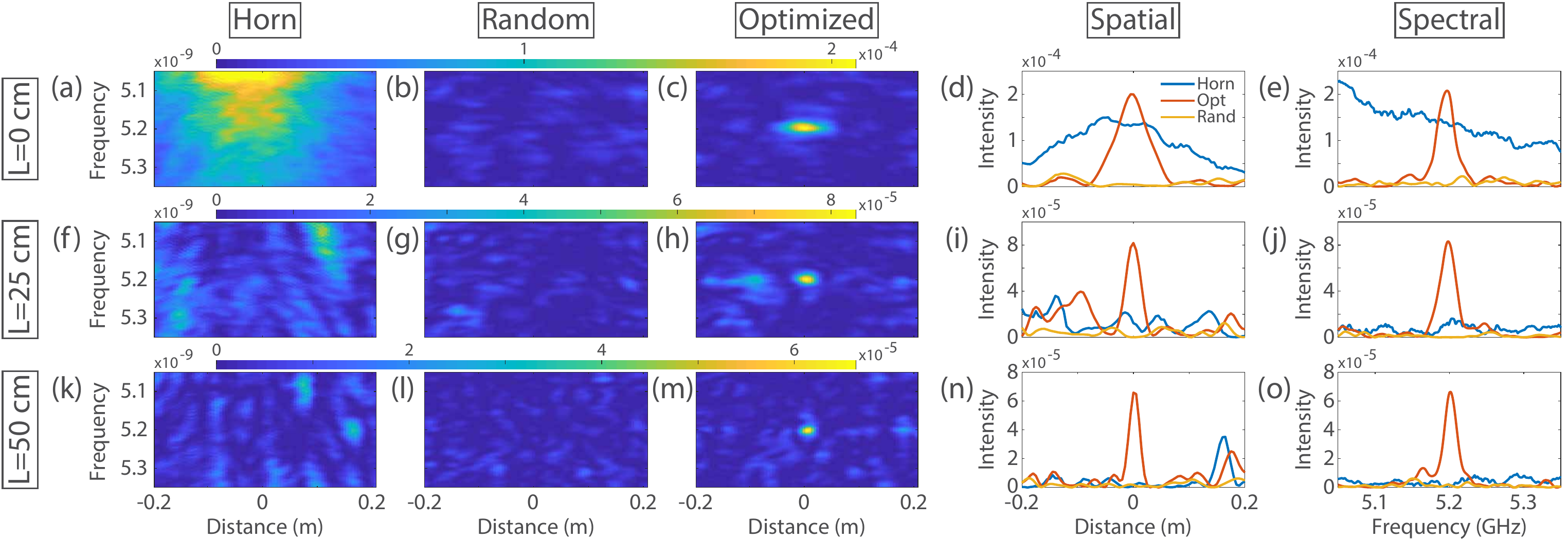}
    \caption{
    Spatial and spectral representation of the transmitted intensity for different configurations of scattering panels: $L=0$~cm (top row), $L=25$~cm (middle row), and $L=50$~cm (bottom row). (a-c),(f-h), and (k-m) display the color-scale representation of the transmission $T(\nu,y)$ measured with a vector network analyzer between 5.05 and 5.35~GHz. (a),(f), and (k) show the transmission using the horn antenna as the transmitter, while (b),(g), (l) corresponds to the random metasurface state, and (c),(h),(m) to the optimized metasurface states, with the reconfigurable cavity as the transmitter. (d),(i),(n) presents the spatial resonance profiles at the working frequency, $\nu_0=5.2$~GHz, measured with the horn antenna (blue), the optimized (red) and random (yellow) metasurface state with the reconfigurable cavity. (e),(j),(o) show the transmitted spectra using the same color scheme.}
    \label{fig:: 1D Scan Data}
\end{figure*}

We  scan the field transmitted through the scattering medium $t(\nu,y)$ between $5.05$ and $5.35$~GHz using a vertically oriented dipole antenna connected to a vector network analyzer (VNA). The dipole is translated along the $y$-axis over a $40$~cm range in steps of $\Delta y = 0.5$~cm ($\Delta y \sim \lambda/10$). The transmission, $T(\nu,y) = |t(\nu,y)|^2$, representing the spatial distribution of the transmitted intensity as a function of frequency and space  are shown in Fig~\ref{fig:: 1D Scan Data} for different sources and for the three thicknesses of scattering panels. The spatial and spectral distributions of transmitted intensity, $T(\nu_0,y)$ and $T(\nu,y=0)$ are also represented in Fig~\ref{fig:: 1D Scan Data}(d-o).


For a horn antenna in the absence of a scattering medium ($L = 0$~cm), the directional beam in the azimuthal plane ($\sim 20^{\circ}$) produces a main lobe with a Full Width at Half Maximum (FWHM) of around 26.5~cm. However, when a scattering medium is introduced between the source and the receiving antenna, multiple scattering within the medium gives rise to a speckle pattern characterized by random fluctuations in both space and frequency. Transmission is strongly attenuated, an effect that becomes increasingly pronounced with greater medium thickness $L$. For uncontrolled illumination, such as a plane wave or, here, an aperture antenna, the average transmission through a diffusive medium characterized by a mean free path $\ell$ indeed scales as $\ell/L$.

For the leaky cavity with random configurations of the metasurfaces, the transmission is further reduced due to two compounding effects: the random speckle pattern radiated outside the cavity and intrinsic absorption within the cavity. We estimate the average intensity radiated through the aperture as

\begin{equation}
    \langle |E(x,y)|^2 \rangle = E_0^2  \frac{\tau_{\mathrm{ap}}^{-1}}{\tau_{\mathrm{ap}}^{-1}+\tau_{\mathrm{abs}}^{-1}},  
\end{equation}

\noindent where $\tau_\mathrm{ap}$ and $\tau_\mathrm{abs}$ represent the leakage time through the aperture and the cumulative intrinsic absorption time of the cavity and the metasurfaces, respectively. $E_0^2$ corresponds to the ideal situation in which all the energy is dissipated through the aperture, i.e., $\tau_{\mathrm{abs}} \rightarrow \infty$. We first estimate the total reverberation time, $\tau^{-1} = \tau_{\mathrm{abs}}^{-1} + \tau_{\mathrm{ap}}^{-1}$, from measurements of the reflection coefficient inside the cavity for $200$ random configurations of the metasurfaces. This yields $\tau = 12.5$~ns, corresponding to a quality factor $Q = 2\pi \nu_0 \tau = 410$ at the central frequency $\nu_0 = 5.2$~GHz. We then close the cavity to suppress radiation through the aperture. In this case, the reverberation time depends only on absorption, giving $\tau = \tau_{\mathrm{abs}} = 16.6$~ns. From these two measurements, we estimate $\tau_{\mathrm{ap}} = 50.6$~ns. The relative fraction of energy leaking through the aperture is therefore $\tau_{\mathrm{ap}}^{-1} / (\tau_{\mathrm{ap}}^{-1}+\tau_{\mathrm{abs}}^{-1}) = 0.25$, indicating that $75$\% of the energy is absorbed within the cavity. Such significant absorption is not negligible and is expected to deteriorate the achievable focusing quality.

We now optimize the metasurface configuration to achieve focusing at the target located at position $\mathbf{r}_0$ and at frequency $\nu_0$. This is achieved through an iterative optimization procedure using the measured intensity as feedback. We start from the best transmitted intensity within a hundred of random configurations and then sequentially adjust the state of each programmable pixel to maximize the intensity at the focal point. While this method effectively enhances focusing, it does not ensure a globally optimal configuration; superior solutions may exist that are not reached through this local optimization approach. The field radiated by the leaky cavity can be expressed in terms of the field radiated through the aperture $E(\mathbf{r}_a)$ and the transmission coefficients between the aperture and the probe location $\mathbf{r}$ as

\begin{equation}
    E(\mathbf{r}) = \int_{\mathbf{r}_a} E(\mathbf{r}_a)t(\mathbf{r}_a,\mathbf{r}) d\mathbf{r}_a.   
\end{equation}

\noindent Ideally, the optimization of the metasurfaces would result in a field at the aperture that is the complex conjugate of the normalized transmission coefficient between the focal point $\mathbf{r}$ and the aperture $E_{\mathrm{opt}}(\mathbf{r_a}) \propto t^*(\mathbf{r}_0,\mathbf{r}_a)$~\cite{Vellekoop2008a}.  We quantify the non-optimality of the wavefront at the aperture by the overlap coefficient $\gamma$, 

\begin{equation}
    \gamma = \frac{\int_{\mathbf{r}_a} E(\mathbf{r_a}) t(\mathbf{r}_0,\mathbf{r}_a) d\mathbf{r}_a}{\sqrt{\int_{\mathbf{r}_a} |E(\mathbf{r_a})|^2 d\mathbf{r}_a \int_{\mathbf{r}_a} |t(\mathbf{r}_0,\mathbf{r}_a)|^2 d\mathbf{r}_a}}.
\end{equation}

\noindent An ideal degree of control would give $E(\mathbf{r_a}) = E_{\mathrm{opt}}(\mathbf{r_a})$ and $|\gamma|^2 = 1$. However, it is never possible to perfectly reproduce this optimal wavefront experimentally because one can only \textit{partially} control the field. Within the reconfigurable cavity, unstirred components arise from propagating paths between the source and the aperture that are either not scattered by the metasurfaces or encounter the metasurface but are not stirred due to its nonzero “structural” scattering cross section. These unstirred paths will dramatically lower the degree of control over the field at the aperture and therefore $\gamma$. For a focusing process, the transmitted intensity can then be approximated by

\begin{equation}
    T_{\mathrm{foc}}(\mathbf{r}) = |\gamma|^2 E_0^2 \frac{\tau_{\mathrm{ap}}^{-1}}{\tau_{\mathrm{ap}}^{-1}+\tau_{\mathrm{abs}}^{-1}} \left|\int_{\mathbf{r}_a} t(\mathbf{r},\mathbf{r}_a) t^*(\mathbf{r}_0,\mathbf{r}_a) d\mathbf{r}_a \right|^2.   
\end{equation}

\noindent $T_{\mathrm{foc}}(\mathbf{r})$ therefore directly depends on the integral over the aperture of the field-field correlation function $t(\mathbf{r},\mathbf{r}_a) t^*(\mathbf{r}_0,\mathbf{r}_a)$.

In absence of a scattering medium ($L=0$~cm), the optimization yields a well-defined focus at the target with a FWHM of $\delta y = 7$~cm, Fig~\ref{fig:: 1D Scan Data}(d). This result aligns with the theoretical formula $\delta y = \lambda F / D \sim 6.9$~cm, where $F \sim 60$~cm is the focal distance and $D = 50$~cm is the aperture of the leaky cavity. The reduced FWHM compared to that of a horn antenna indicates improved spatial focusing and increased intensity at the focal point. However, the overall intensity enhancement remains limited, with an enhancement factor of only 1.4,  primarily due to losses within the cavity. The enhancement factor between the optimized and the direct horn-to-dipole reference intensity increases drastically in the presence of the scattering medium, reaching a factor of about 50 for $L=50$~cm, Fig.~\ref{fig:: 1D Scan Data}(m-o). For a diffusive medium, the waves are focused on the scale of $\lambda/2$, which corresponds to the width of the field-field correlation function $|\int_{\mathbf{r}_a} t(\mathbf{r},\mathbf{r}_a) t^*(\mathbf{r}_0,\mathbf{r}_a) d\mathbf{r}_a |^2$ \cite{vellekoop2010exploiting}. We observe that for $L=50$~cm, the lateral dimension of the focal spot is indeed $\delta y \sim 2$~cm, which is approximately $\lambda /2 = 2.8$~cm. 

\begin{figure}
    \centering
    \includegraphics[width=8.5cm]{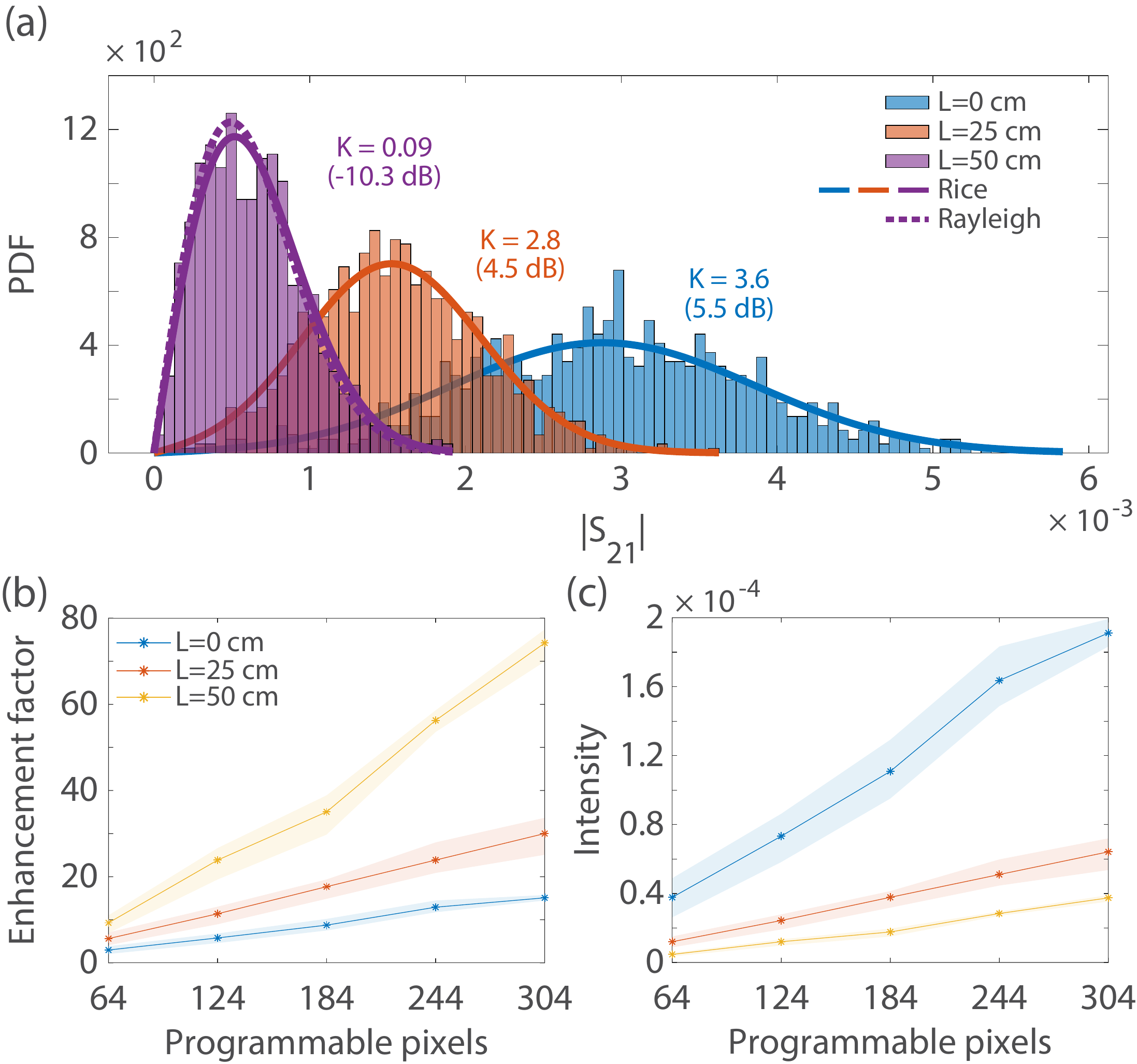}
    \caption{(a) Probability density functions (PDFs) of the field amplitude $|E|$ over $m=900$ random metasurface configurations, for $L=0$~cm (blue), $L=25$~cm (orange) and $L=50$~cm (magenta). Solid and dotted lines represents Rician and Rayleigh distributions respectively, based on experimentally extracted $\sigma$ and $\mu$ parameters.
    (b) Enhancement factor $\eta = T_\mathrm{foc} / \langle T \rangle$ as a function of the number of controlled pixels $N_\mathrm{p}$ for each value of $L$.
    (c) Intensity of the field at the focal point, $T_\mathrm{foc}$, for the same configurations.
    }
    \label{fig:: Rician - Sweep Npixt}
\end{figure}

We now investigate the impact of the scattering medium on unstirred paths and, therefore, on the parameter $\gamma$. For a random configuration of the metasurface pixels (i.e., in absence of optimization), the field at the probe location $\mathbf{r}$ can be decomposed as $E(\mathbf{r}) = E_u(\mathbf{r}) + \delta E(\mathbf{r})$, where $E_u(\mathbf{r})$ is the unstirred component, independent of the metasurface configuration, and $\delta E(\mathbf{r})$ is the stirred contribution fluctuating with the metasurface states. $\delta E(\mathbf{r})$ is typically modeled as a zero-mean complex random Gaussian variable, i.e., $\langle \delta E(\mathbf{r}) \rangle = 0$, where $\langle \cdots \rangle$ denotes averaging over random configurations of the metasurfaces. As a result, the statistical distribution of the field amplitude $|E(\mathbf{r})|$ is characterized by a Rician distribution.
In the limiting case where the stirred component dominates, i.e., $E(\mathbf{r}) \simeq \delta E(\mathbf{r})$, the distribution reduces to a Rayleigh distribution. 
We define the ratio of the unstirred and stirred contributions by the parameter $\kappa =  |E_u(\mathbf{r})|^2 / \langle |\delta E(\mathbf{r})|^2 \rangle$, which corresponds to the Rician \textit{K}-factor for an infinite number of random configurations. 
This factor, commonly used in the context of reverberation chambers (RC)~\cite{chen2011estimation,remley2016configuring}, 
can be experimentally estimated from the statistical distribution of the complex transmitted field measured over $m$ random metasurface configurations:~\cite{lemoine2010k,ahmed2024over}

\begin{equation}
    K = \frac{m-2}{m-1} \frac{\mu^2}{2\sigma^2} - \frac{1}{m},
\end{equation}

\noindent where $\mu^2$ is the squared modulus of the mean of the transmission coefficients, and $\sigma^2$ denotes their variance. Since these coefficients are complex-valued, we have $2\sigma^2 = \mathrm{Var}(E) = \mathrm{Var}(\Re(E)) + \mathrm{Var}(\Im(E))$, with $E$ being the $m$ measured complex transmission coefficients. 

The measured \textit{K}-factors for the three different configurations of the scattering medium for $m=900$ using all pixels are: $\textit{K} = 3.6$ (5.5~dB) for $L = 0$~cm, $\textit{K} = 2.8$ (4.5~dB) for $L = 25$~cm, and $\textit{K} = 0.09$ (-10.3~dB) for $L = 50$~cm. These results show that the presence of the scattering medium enhances the proportion of the stirred component, increasing the degree of control exerted by the metasurfaces. Backscattering on the diffusive medium generates new scattering paths within the cavity and, therefore, strongly enhances the stirring process. In particular, the extremely low value of $K$ for $L = 50$~cm indicates that the field becomes almost entirely controlled across configurations. Figure~\ref{fig:: Rician - Sweep Npixt}(a) shows the probability density functions for these three cases. As expected, each follows a Rician distribution (solid lines); however, for $L=50$~cm, the distribution closely approximates a Rayleigh distribution (dotted line), confirming the negligible contribution of the unstirred field component in this case. Note that the Rician and the Rayleigh distributions are not fits but are based on the experimentally extracted parameters $\sigma$ and $\mu$.

To confirm experimentally the enhancement of the degree of control over the transmitted field in the presence of a strongly scattering medium, we measure the focused intensity and the enhancement factor $\eta = {T_\mathrm{foc}}/{\langle T \rangle}$ as a function of the number of controlled programmable pixels of the metasurfaces $N_\mathrm{p}$. To this end, for each value of $N_\mathrm{p} \in \{64,124,184,244,304\}$ involved during the optimization process, five independent optimizations are performed, yielding five distinct optimized states corresponding to different focal intensities $T_\mathrm{foc}$. 900 random configurations are used to obtain $\langle T \rangle$. The averaged enhancement factors $\eta$ over these five realizations are shown as solid lines in Fig.~\ref{fig:: Rician - Sweep Npixt}(b). The shaded areas represent the full variation range of measured values. Although the optimization process does not guarantee an optimal solution, the intensities measured across the five independent realizations remain closely clustered. This consistency suggests a strong robustness in the optimization process.

We observe a clear linear increase in the enhancement factor $\eta$ with $N_\mathrm{p}$, following the relation $\eta = |\gamma|^2 N_\mathrm{p}$. This is similar to wavefront shaping experiments in optics through opaque samples, for which $\eta$ increases as $|\gamma|^2 N$, where $N$ is the number of controlled pixels on a spatial light modulator \cite{Vellekoop2008a}. In our case, most importantly, the coefficient $|\gamma|^2$ estimated from a linear fit of $\eta$ with $N_\mathrm{p}$ depends on $L$, with $|\gamma|^2 \sim 0.05$ for $L=0$~cm, $|\gamma|^2 = 0.09$ for $L=25$~cm and $|\gamma|^2 = 0.26$ for $L=50$~cm. This trend correlates with the measured decrease in the Rician K-factors, indicating a stronger contribution from the dynamically varying (stirred) component of the field and an improved degree of control over the radiated field. Consequently, this leads to a greater enhancement of the intensity at the focal point relative to the horn antenna reference. While the enhancement is relatively modest in the absence of the scattering medium, it increases significantly with medium thickness, reaching a value of 74 for $L=50$~cm. 

These findings further support the conclusion that the presence of the scattering medium enhances system controllability by suppressing the contribution of the unstirred component. In this regime, the field radiated through the aperture at location $\mathbf{r}$ becomes highly sensitive to the metasurface configuration, enabling efficient and effective wavefront shaping through optimization.

 
\section{Wireless communication through the scattering medium}
\label{sec:trans_xp} 

We now implement a realistic communication channel through a scattering medium for $L=50$~cm. The signals are emitted and measured through two software-defined radios (SDRs) enabling the transmission and reception of arbitrary data and time windows. The receiving SDR is connected to the transition inside the cavity, while the transmitting one is connected to the simple wire dipole visible in Fig~\ref{fig:: Experimental SetUp}. Our goal here is to transmit a picture from the transmitter behind the scattering medium to the inside of the cavity and verify how the cavity can be tuned to perceive low power transmission.

First,  the transmitter emits a continuous carrier wave with a fixed power level of $g~\mathrm{dBm}$, and the metasurface configuration is optimized to maximize the intensity at the receiver. The information is not transmitted at this stage. This controlled transmission ensures measurement repeatability and consistent evaluation of metasurface configurations. In practical scenarios where direct control of the transmitter is not achievable, such optimization could alternatively be based on averaging over many symbols or by detecting a known, constant-power preamble embedded in the transmission protocol. Second, the transmitter sends an amplitude modulated picture through the channel, sending each RGB mask sequentially, line by line, pixel by pixel. Each of the 255 possible pixel intensity values is mapped to a specific amplitude of the carrier signal, with a maximum transmission power of $g$~dBm. These modulated signals are received by the optimized cavity, decoded based on their amplitude levels, and subsequently rasterized to reconstruct the transmitted image.

\begin{figure}
    \centering
    \includegraphics[width=8.5cm]{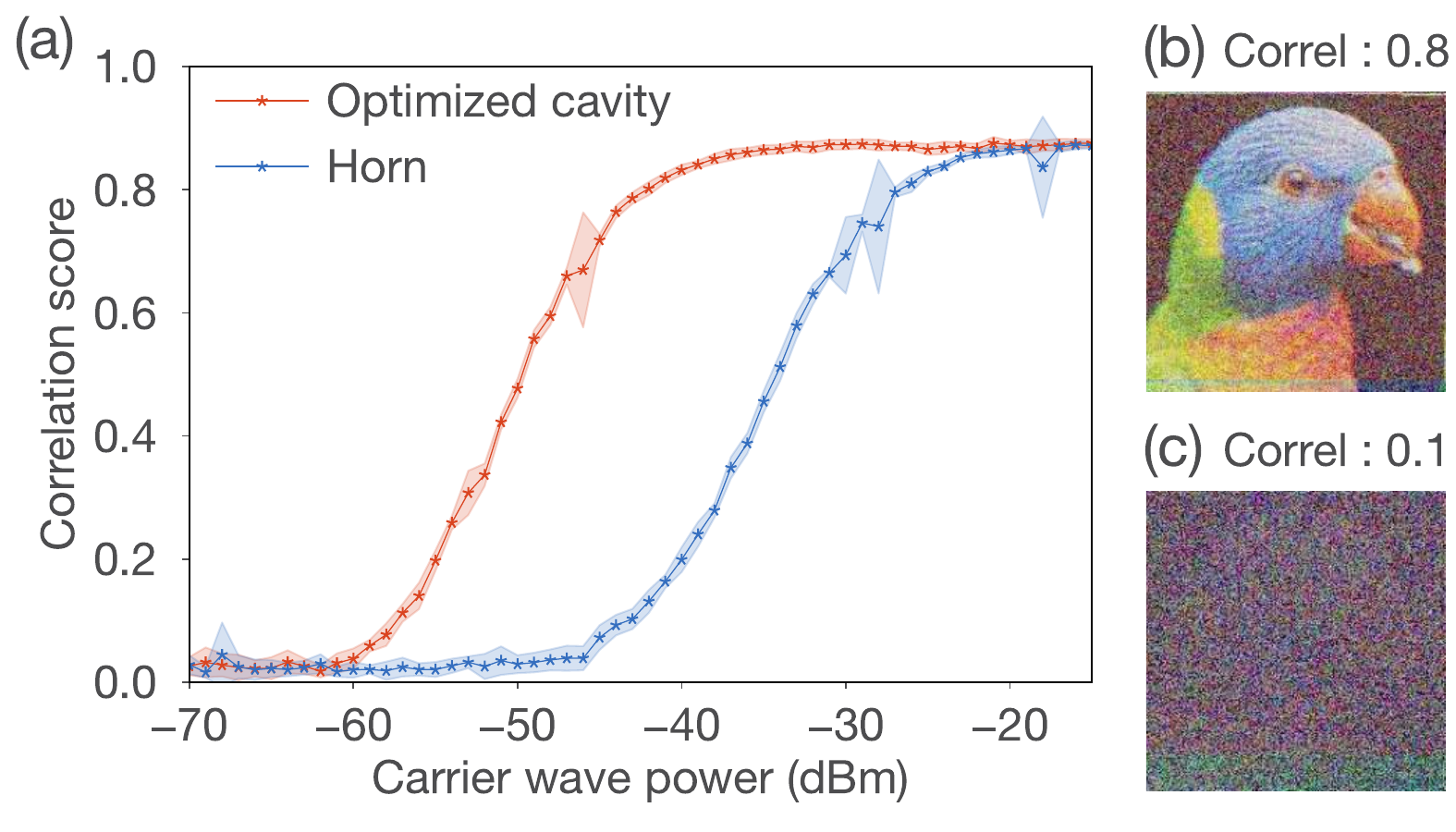}
    \caption{(a) Correlation score as a function of incident power level $g$ for an optimized cavity (red) and a horn antenna (blue). The cavity is iteratively optimized for each incident power. The error bar is computed from 10 transmissions of $200 \times 200~\mathrm{px}$ pictures. (b,c) Example of recovered pictures for a power level $g = -45~\mathrm{dBm}$ for an optimized cavity (b) and a horn antenna (c). The corresponding correlation scores are approximately $0.8$ and $0.1$, respectively.}
    \label{fig::trans_single_compose}
\end{figure}

To evaluate the channel performance across varying transmission powers, we repeat the previous steps for different incident powers $g$ between $-70$ and $-15~\mathrm{dBm}$. To ensure that low incident power doesn't hinder the optimization process,  the metasurface optimization is re-executed for each $g$, before estimating the degree of correlation between the transmitted and received images. The results are presented in Fig~\ref{fig::trans_single_compose}(a,b,c) and compared to the case of a horn antenna acting as a receiver. Due to ambient noise and imperfections in the rastering process, the maximum correlation score achieved was approximately equal to $0.9$. Within the context of this experiment, this value will be considered as optimal. Overall, the optimized cavity-based channel demonstrates a performance gain of approximately $15~\mathrm{dB}$ over the horn antenna using standard off-the-shelf SDR, even in low-power scenarios. 
 
We emphasize that even in the case of a small incident power, we were still able to optimize the metasurfaces within the leaky cavity. For instance, at $g = -50~\mathrm{dBm}$ in Fig~\ref{fig::trans_single_compose}, while the horn is not capable of perceiving any information from the transmitted picture, a starting point for the iterative optimization process can be found, ultimately converging to a near-optimal configuration.

 
\section{Backscatter experiments}
\label{sec:back_xp} 

\begin{figure}
    \centering
    \includegraphics[width=8.5cm]{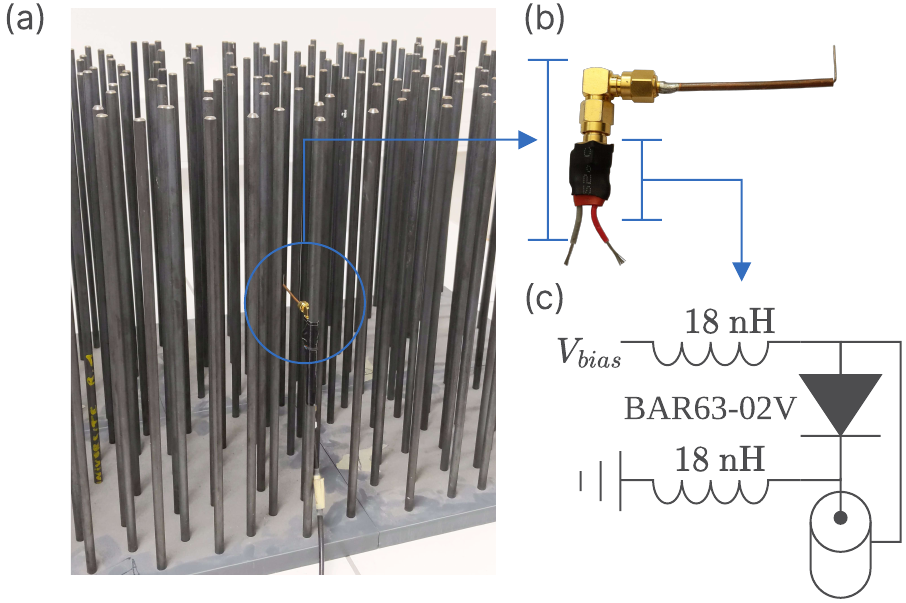}
    \caption{(a) Scattering panels with the added reflector placed at mid height, $z\simeq25$~cm, of a scattering panel rod near the edge.
    Backscattering element load (b) and it's schematic (c).}
    \label{fig::schematic}
\end{figure}

We finally demonstrate that a programmable leaky cavity can be employed for backscattering communication through a random scattering medium. In contrast to the previous configuration, which utilized a conventional transmitting antenna, we now replace the antenna with a passive backscattering element that encodes data by modulating its reflectivity. In this configuration, the leaky cavity plays a dual role: it both illuminates the backscatterer and captures the modulated signal reflected from it.

Measurements are carried out using a VNA instead of a SDR. The choice of the VNA is motivated by its significantly higher dynamic range ($130~\mathrm{dB}$), which is essential for accurately characterizing and optimizing the metasurface response. In contrast, the limited dynamic range of SDRs does not provide sufficient sensitivity for precise wavefront control in the presence of scattering and weak backscattered signals. Another limitation associated with SDRs is their inability to transmit and receive through the same port simultaneously. This constraint necessitates two separate transitions within the cavity—one for transmission and one for reception—to create a multipath scenario. However, this configuration diminishes the benefits of cavity-based optimization, as it introduces additional complexity and potential signal degradation. Moreover, having distinct transmit and receive paths in close proximity can lead to receiver saturation, where the receiver becomes easily blinded by the strong nearby illumination signal, further compromising sensitivity and communication performance.

The backscattering device is placed inside the scattering medium, near its outermost edge, at a distance $L = 45$~cm (see Fig.\ref{fig::schematic}(a)). It consists of a dipole antenna with an impedance $Z_{\mathrm{ant}}$ connected to a variable load at its feed point. As illustrated in Fig.\ref{fig::schematic}(b,c), the variable load comprises a diode placed between the inner conductor and outer shield of an SMA connector, with two $18\mathrm{nH}$ inductors acting as RF blockers to prevent incoming signals from coupling into the biasing circuit. A bias voltage $V_\mathrm{bias}$ is supplied using a waveform generator, enabling dynamic control of the load's impedance and, consequently, its reflectivity. When $V_\mathrm{bias} = 0~\mathrm{V}$, the diode approximates a short circuit with an impedance $Z_{\mathrm{short}}$, whereas at $V_\mathrm{bias} = 1~\mathrm{V}$ it approaches an open circuit with an impedance $Z_{\mathrm{open}}$.  The reflection coefficient $\Gamma$ corresponding to these two states is given by:
 
\begin{equation}
\Gamma = \frac{Z_{\mathrm{[short,open]}} - Z^{*}_{\mathrm{ant}}}{Z_{\mathrm{[short,open]}} + Z_{\mathrm{ant}}} 
\end{equation}  
 
\noindent Since neither $Z_{\mathrm{short}}$ nor $Z_{\mathrm{open}}$ constitutes a perfect short, open, or matched condition, different reflection changes will be created in both phase and amplitude. These variations enable discrimination between backscatter states during interrogation. 

To transmit an image through the backscattering-based communication channel, data are encoded using a binary modulation scheme applied to the device, with two distinct states: a 0 corresponds to the short-circuit state, while a 1 represents the open-circuit state. Although an analogical encoding of the 255 distinct states using $V_\mathrm{bias}$ between $0$ and $1~V$ would be theoretically feasible, the presence of the scattering medium significantly limits the measurement dynamic range. In such realistic conditions, smaller impedance variations cannot be reliably detected. The task, therefore, reduces to estimating the backscattering state from $S_{11}$ measurements, where $S_{11}^{\mathrm{open}}$ and $S_{11}^{\mathrm{short}}$ denote the reflection coefficients corresponding to the open and short states, respectively.

Before evaluating the channel reliability across the illumination power range $g \in [-40,10]~\mathrm{dBm}$, we first optimize the cavity for $g = 10~\mathrm{dBm}$. As the same port within the leaky cavity is used for both transmission and reception, directly maximizing the received signal intensity would primarily enhance internal reflections rather than the desired interaction with the backscattering device. To avoid this, the optimization procedure involves alternating the device between its two states for each metasurface configuration. The metasurfaces are then optimized to maximize the following cost function:

\begin{equation}
    \mathcal{F} = \left||S_{11}^{\mathrm{open}}| - |S_{11}^{\mathrm{short}}|\right|.
\end{equation} 

 
\noindent This criterion promotes configurations that maximize the contrast between the two backscatter states in terms of the reflected signal amplitude. 


\begin{figure}
    \centering
    \includegraphics[width=8.5cm]{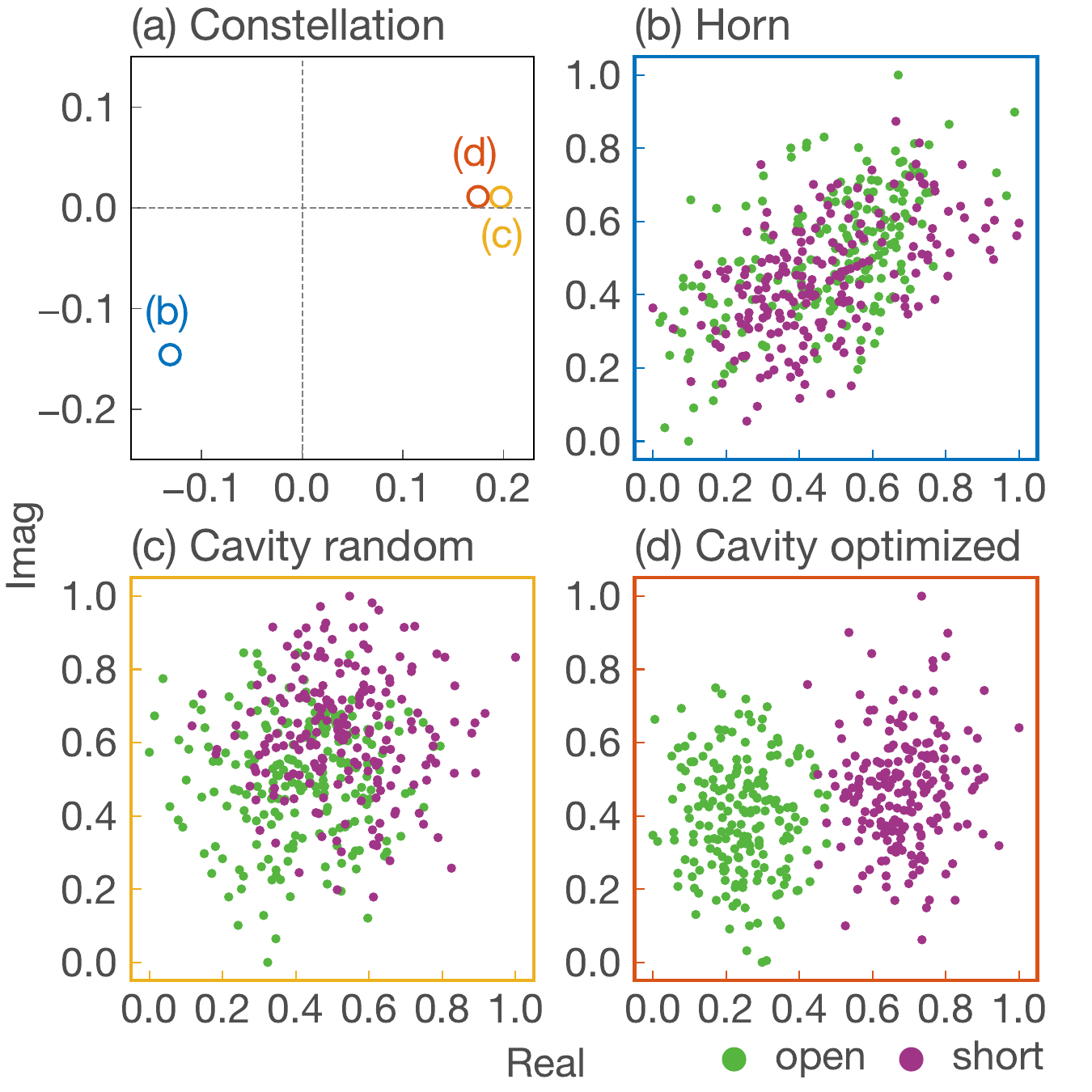}
    \caption{ (a) Average reflection parameter $\langle S_{11} \rangle$ measured  at $g=0~\mathrm{dBm}$ shown in the complex plane (real and imaginary components) for a source being a horn antenna (blue), a random configuration (orange) and an optimal cavity (red) configuration. (b,c,d) Clouds of $S_{11}$ for the two states of the modulated backscatter for the horn antenna (b), a random cavity configuration (c) and an optimized configuration of the cavity (d). Each plot is made of 200 points with the backscatter in an open state (green dots) and 200 points in a closed state (magenta dots). Points were normalized to occupy the same area of the complex plane. 
    }
    \label{fig::constel}
\end{figure}

The reflection coefficient distributions in the complex plane, shown in Fig.~\ref{fig::constel}(a,b,c,d), illustrate the impact of noise on state estimation precision. In the case of an unoptimized cavity, the clusters corresponding to the two backscatter states significantly overlap (Fig.~\ref{fig::constel}(c)), making it impossible to distinguish between them and, consequently, to perform any meaningful optimization. To enhance the signal-to-noise ratio during the optimization process, we average over 100 acquisitions for each configuration. This averaging allows the two reflection clusters to become well-separated by the end of the optimization (Fig.~\ref{fig::constel}(d)), enabling reliable state identification from a single measurement. For comparison, the same procedure is applied using a horn antenna positioned in front of the sample, aligned with the backscattering device. In this configuration, the separation between the cluster centers is smaller than their spread, resulting in a low probability of correctly identifying the backscatter state (Fig.~\ref{fig::constel}(b)).

\begin{figure}
    \centering
    \includegraphics[width=8.5cm]{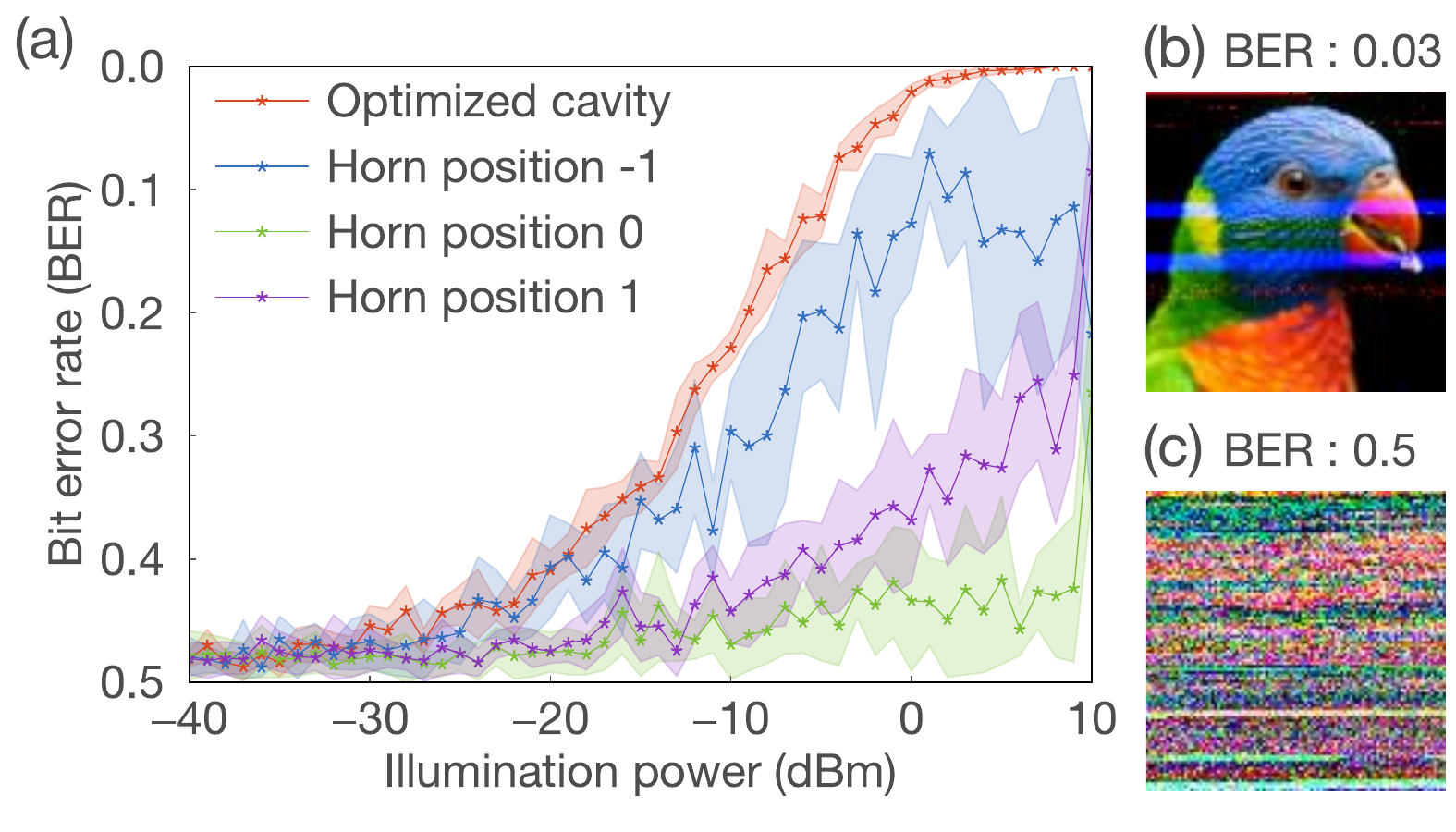}
    \caption{Measurement of the communication channel performance depending on the illumination power. 
    (a) Evolution of the Bit error rate (BER) as a function of the illumination power of the cavity feed (red) or horn antenna feed for three different positions (blue, green, magenta). The cavity is optimized only once at $g=0~\mathrm{dBm}$, each point and error is based on 2000 bits. 
    (b),(c) show an example of recovered picture at BER score of approximately $0.03$ or $0.5$.
    The horn antenna positions are varied along the y-axis with  $2\lambda~ (11.4~\mathrm{cm})$  steps.
    }
    \label{fig::back_single_compose}
\end{figure}

After optimization, we transmit an image through the backscattering-based communication channel to assess its reliability. This is quantified by measuring the bit error rate (BER) over 2000 transmitted bits for each illumination power level $g$. Across the 2000 transmitted bits, an equal number of 0s and 1s were sent. Each measurement obtained via the cavity was labeled according to the true backscatter state. Classification of each measurement as a 0 or 1 was performed based on whether its amplitude was below or above the mean amplitude of all 2000 measurements. A proper cavity optimization should result in a distribution similar to that shown in Fig.~\ref{fig::constel}(d), where the clusters corresponding to the two states are sufficiently separated. The BER is then calculated as the fraction of misclassified bits relative to the ground-truth labels.

At low VNA power, specifically below $-30~\mathrm{dBm}$, noise-induced fluctuations in the reflection parameters prevent reliable discrimination between the two signal states. As a result, the probability of correctly or incorrectly identifying a bit becomes equal, yielding a BER of 0.5. As the incident power increases, the impact of noise fluctuations is progressively reduced, leading to improved state discrimination. For $g>0~\mathrm{dBm}$, the two signal states can be distinguished with near-perfect accuracy, and the BER approaches $0$. Figure~\ref{fig::back_single_compose}(b) shows the image transmitted at an incident power of $0~\mathrm{dBm}$, corresponding to a BER of $0.03$. The reconstructed image closely matches the original, with the exception of two horizontal blue lines. These artifacts are attributed to a modification of the environment occurring in the vicinity of the channel. Since the acquisition of all samples associated with a single image takes a few hours, any physical change (such as the movement of people in the experiment room) during this period can distort the perceived backscatter reflection. Incorporating time-gating could mitigate these effects at the expense of reduced acquisition speed. Despite minor visual artifacts, the small focal spot produced by the optimized cavity enhances the noise robustness of the measurements. This improvement is primarily evidenced in the narrower error bars of the optimized cavity's correlation values in Fig.~\ref{fig::back_single_compose}(a).

We now compare these results with those obtained using a horn antenna placed in front of the scattering medium, replacing the leaky reconfigurable cavity. Due to strong spatial intensity fluctuations induced by the scattering medium, a single horn antenna measurement does not provide a reliable estimate of the average BER. To address this, we conduct BER measurements at three different static positions of the horn antenna, see  Fig.~\ref{fig::back_single_compose}(a). While the case of the horn antenna consistently underperforms compared to an optimized cavity, its performance varies significantly with position. Notably, these variations persist even when the backscattering element lies within the horn's main radiation lobe. This observation indicates that performance is dominated by the complex wave interference and scattering effects within the medium, rather than by the simple line-of-sight alignment or directivity of the source.

\section{Conclusion}
In this work, we have demonstrated the effectiveness of a programmable leaky cavity for enhancing wave transmission and information retrieval through disordered scattering media. By leveraging reconfigurable metasurfaces, the cavity allows precise wave focusing, resulting in a significant increase in the focused intensity at a target location, even in the presence of strong multiple scattering. We have even shown that the degree of control over the transmitted wavefront is enhanced by the presence of the scattering medium because unstirred field contributions are suppressed as a result of backscattering within the cavity. We have further shown that this approach not only improves conventional wave focusing but also enables robust wireless communication through complex media. Finally, we have extended its application to a backscatter communication scenario, highlighting that the same reconfigurable cavity can be used to illuminate and interrogate a passive, impedance-modulating device through a scattering layer.

Beyond backscatter communication, the programmable leaky cavity approach holds significant potential in several other domains. In particular, it may help address one of the longstanding challenges in wireless power transfer: efficiently focusing electromagnetic energy through complex environments, such as walls.
By optimizing the wavefront, the cavity can enhance energy delivery to specific locations with minimal leakage, thereby improving both efficiency and safety. Similarly, in non-destructive through-the-wall imaging, the ability to focus waves opens up possibilities where traditional techniques often fail due to scattering-induced signal loss. 

Finally, we highlight that our approach may be of great interest in electromagnetic cybersecurity. Radio-frequency retroreflector attacks involve interrogating a modulated hardware Trojan from the far field. In such scenarios, an antenna is coupled to a transmission line carrying data—such as a VGA video signal or a USB keyboard input—and the voltage variations modulate the antenna’s impedance via a connected transistor~\cite{Kinugawa_Fujimoto_Hayashi_2019, Wakabayashi_usenix_rfra_2018, granier:hal-05247457}. Much like in backscatter communication, the attacker’s goal is to retrieve the Trojan’s state remotely. However, these attacks are fundamentally limited by a low signal-to-noise ratio, particularly in through-the-wall or cluttered environments where multiple scattering dominates. Our approach, leveraging a programmable leaky cavity, could significantly improve the viability and range of such attacks, offering a new pathway for long-distance, low-power Trojan interrogation under real-world conditions.

\section*{Acknowledgements}

This work is supported by the European Union through the European Regional Development Fund (ERDF), Ministry of Higher Education and Research, CNRS, Brittany region, Conseils D{\'e}partementaux d’Ille-et-Vilaine and C{\^o}tes d’Armor, Rennes M{\'e}tropole, and Lannion Tr{\'e}gor Communaut{\'e}, through the CPER Project CyMoCod, and by the French “Agence Nationale de la Recherche” (ANR) under the Grant ANR-24-CE91-0007-01 for the project META-INCOME.


\bibliographystyle{apsrev4-1}

\bibliography{bibliography.bib}

@PREAMBLE{
 "\providecommand{\noopsort}[1]{}" 
 # "\providecommand{\singleletter}[1]{#1}%" 
}

@inproceedings {Wakabayashi_usenix_rfra_2018,
   author = {Satohiro Wakabayashi and Seita Maruyama and Tatsuya Mori and Shigeki Goto and Masahiro Kinugawa and Yuichi Hayashi and Michael Smith},
   title = {A Feasibility Study of Radio-frequency Retroreflector Attack},
   booktitle = {12th USENIX Workshop on Offensive Technologies (WOOT 18)},
   year = {2018},
   address = {Baltimore, MD},
   url = {https://www.usenix.org/conference/woot18/presentation/wakabayashi},
   publisher = {USENIX Association},
   month = aug
}

@article{Kinugawa_Fujimoto_Hayashi_2019, 
   title={Electromagnetic Information Extortion from Electronic Devices Using Interceptor and Its Countermeasure}, 
   volume={2019},  url={https://tches.iacr.org/index.php/TCHES/article/view/8345},  
   DOI={10.13154/tches.v2019.i4.62-90}, 
   abstractNote={The problem of information leakage through electromagnetic waves for various devices has been extensively discussed in literature. Conventionally, devices that are under such a threat suffer from potential electromagnetic information leakage during their operation. Further, the information inside the devices can be obtained by monitoring the electromagnetic waves leaking at the boundaries of the devices. The leakage of electromagnetic waves, however, was not observed for some devices, and such devices were not the target of the threat discussed above. In light of this circumstance, this paper discusses an “interceptor” that forces the leakage of information through electromagnetic waves, even from devices in which potential electromagnetic leakage does not occur. The proposed interceptor is a small circuit consisting of an affordable semiconductor chip and wiring and is powered by electromagnetic waves that irradiate from the outside of a device as its driving energy. The distance at which information is obtained is controlled by increasing the intensity of the irradiated electromagnetic waves. The paper presents the structure of the circuit for implementing the proposed interceptor to be used in major input–output devices and cryptographic modules, mounting a pathway designed on the basis of the construction method onto each device. Moreover, it is shown that it is possible to forcefully cause information leakage through electromagnetic waves. To detect the aforementioned threat, the paper also focuses on the changes in a device itself and the surrounding electromagnetic environment as a result of mounting an interceptor and considers a method of detecting an interceptor by both passive and active monitoring methods.}, 
   number={4}, 
   journal={IACR Transactions on Cryptographic Hardware and Embedded Systems}, 
   author={Kinugawa, Masahiro and Fujimoto, Daisuke and Hayashi, Yuichi}, 
   year={2019}, 
   month={Aug.}, 
   pages={62–90} 
}

@inproceedings{granier:hal-05247457,
  TITLE = {{Diode-Based Multi-Trojan RF Retroreflector Attack}},
  AUTHOR = {Granier, Pierre and Nicolas, Marie-A{\"i}nhoa and Lorandel, Jordane and Moy, Christophe and Besnier, Philippe and Davy, Matthieu and Sarrazin, Fran{\c c}ois},
  URL = {https://hal.science/hal-05247457},
  BOOKTITLE = {{EMC Europe 2025 conference Proceedings}},
  ADDRESS = {Paris, France},
  YEAR = {2025},
  MONTH = Sep,
  HAL_ID = {hal-05247457},
  HAL_VERSION = {v1},
}

@article{rotter2017light,
  title={Light fields in complex media: Mesoscopic scattering meets wave control},
  author={Rotter, Stefan and Gigan, Sylvain},
  journal={Rev. Mod. Phys.},
  volume={89},
  number={1},
  pages={015005},
  year={2017},
  publisher={APS}
}

@article{Mosk2012,
   author = {Mosk, Allard P and Lagendijk, Ad and Lerosey, Geoffroy and Fink, Mathias},
   title = {Controlling waves in space and time for imaging and focusing in complex media},
   journal = {Nat. Photonics},
   volume = {6},
   number = {5},
   pages = {283-292},
   year = {2012}
}

@article{vellekoop2007focusing,
  title={Focusing coherent light through opaque strongly scattering media},
  author={Vellekoop, Ivo M and Mosk, Allard P},
  journal={Optics letters},
  volume={32},
  number={16},
  pages={2309--2311},
  year={2007},
  publisher={Optica Publishing Group}
}

@article{Vellekoop2008a,
   author = {Vellekoop, I. M. and Mosk, A. P.},
   title = {Universal Optimal Transmission of Light Through Disordered Materials},
   journal = {Phys. Rev. Lett.},
   volume = {101},
   number = {12},
   pages = {120601-4},
   year = {2008}
}

@article{vellekoop2010exploiting,
  title={Exploiting disorder for perfect focusing},
  author={Vellekoop, Ivo M and Lagendijk, Aart and Mosk, AP},
  journal={Nature photonics},
  volume={4},
  number={5},
  pages={320--322},
  year={2010},
  publisher={Nature Publishing Group UK London}
}

@article{Gerardin2014,
   author = {G{\'e}rardin, Benoît and Laurent, Jérôme and Derode, Arnaud and Prada, Claire and Aubry, Alexandre},
   title = {Full Transmission and Reflection of Waves Propagating through a Maze of Disorder},
   journal = {Phys. Rev. Lett.},
   volume = {113},
   number = {17},
   pages = {173901},
   year = {2014}
}

@article{cao2022shaping,
  title={Shaping the propagation of light in complex media},
  author={Cao, Hui and Mosk, Allard Pieter and Rotter, Stefan},
  journal={Nature Physics},
  volume={18},
  number={9},
  pages={994--1007},
  year={2022},
  publisher={Nature Publishing Group UK London}
}

@article{derode1995robust,
  title={Robust acoustic time reversal with high-order multiple scattering},
  author={Derode, Arnaud and Roux, Philippe and Fink, Mathias},
  journal={Physical review letters},
  volume={75},
  number={23},
  pages={4206},
  year={1995},
  publisher={APS}
}

@article{bender2022depth,
  title={Depth-targeted energy delivery deep inside scattering media},
  author={Bender, Nicholas and Yamilov, Alexey and Goetschy, Arthur and Y{\i}lmaz, Hasan and Hsu, Chia Wei and Cao, Hui},
  journal={Nature Physics},
  volume={18},
  number={3},
  pages={309--315},
  year={2022},
  publisher={Nature Publishing Group UK London}
}

@article{jeong2018focusing,
  title={Focusing of light energy inside a scattering medium by controlling the time-gated multiple light scattering},
  author={Jeong, Seungwon and Lee, Ye-Ryoung and Choi, Wonjun and Kang, Sungsam and Hong, Jin Hee and Park, Jin-Sung and Lim, Yong-Sik and Park, Hong-Gyu and Choi, Wonshik},
  journal={Nature Photonics},
  volume={12},
  number={5},
  pages={277--283},
  year={2018},
  publisher={Nature Publishing Group UK London}
}

@article{cheng2014focusing,
  title={Focusing and energy deposition inside random media},
  author={Cheng, Xiaojun and Genack, Azriel Z},
  journal={Optics letters},
  volume={39},
  number={21},
  pages={6324--6327},
  year={2014},
  publisher={Optica Publishing Group}
}

@article{horstmeyer2015guidestar,
  title={Guidestar-assisted wavefront-shaping methods for focusing light into biological tissue},
  author={Horstmeyer, Roarke and Ruan, Haowen and Yang, Changhuei},
  journal={Nature photonics},
  volume={9},
  number={9},
  pages={563--571},
  year={2015},
  publisher={Nature Publishing Group UK London}
}

@article{bouchet2021maximum,
  title={Maximum information states for coherent scattering measurements},
  author={Bouchet, Dorian and Rotter, Stefan and Mosk, Allard P},
  journal={Nature Physics},
  volume={17},
  number={5},
  pages={564--568},
  year={2021},
  publisher={Nature Publishing Group UK London}
}

@article{yeo2022time,
  title={Time Reversal Communications With Channel State Information Estimated From Impedance Modulation at the Receiver},
  author={Yeo, K Brahima and Leconte, Cecile and Del Hougne, Philipp and Besnier, Philippe and Davy, Matthieu},
  journal={IEEE Access},
  volume={10},
  pages={91119--91126},
  year={2022},
  publisher={IEEE}
}

@article{chen2011estimation,
  title={Estimation of average Rician K-factor and average mode bandwidth in loaded reverberation chamber},
  author={Chen, Xiaoming and Kildal, Per-Simon and Lai, Sz-Hau},
  journal={IEEE Antennas and Wireless Propagation Letters},
  volume={10},
  pages={1437--1440},
  year={2011},
  publisher={IEEE}
}

@article{remley2016configuring,
  title={Configuring and verifying reverberation chambers for testing cellular wireless devices},
  author={Remley, Kate A and Dortmans, Jos and Weldon, Catherine and Horansky, Robert D and Meurs, Thomas B and Wang, Chih-Ming and Williams, Dylan F and Holloway, Christopher L and Wilson, Perry F},
  journal={IEEE Transactions on Electromagnetic Compatibility},
  volume={58},
  number={3},
  pages={661--672},
  year={2016},
  publisher={IEEE}
}

@article{ahmed2024over,
  title={Over-the-air emulation of electronically adjustable Rician MIMO channels in a programmable-metasurface-stirred reverberation chamber},
  author={Ahmed, Ismail and Davy, Matthieu and Prod’homme, Hugo and Besnier, Philippe and Del Hougne, Philipp},
  journal={IEEE Transactions on Antennas and Propagation},
  year={2024},
  publisher={IEEE}
}

@article{lemoine2010k,
  title={On the $ K $-factor estimation for Rician channel simulated in reverberation chamber},
  author={Lemoine, Christophe and Amador, Emmanuel and Besnier, Philippe},
  journal={IEEE Transactions on Antennas and Propagation},
  volume={59},
  number={3},
  pages={1003--1012},
  year={2010},
  publisher={IEEE}
}

@article{PhysRevLett.115.017701,
  title = {Wave-Field Shaping in Cavities: Waves Trapped in a Box with Controllable Boundaries},
  author = {Dupr\'e, Matthieu and del Hougne, Philipp and Fink, Mathias and Lemoult, Fabrice and Lerosey, Geoffroy},
  journal = {Phys. Rev. Lett.},
  volume = {115},
  issue = {1},
  pages = {017701},
  numpages = {5},
  year = {2015},
  month = {Jul},
  publisher = {American Physical Society},
  doi = {10.1103/PhysRevLett.115.017701},
  url = {https://link.aps.org/doi/10.1103/PhysRevLett.115.017701}
}

@article{sleasman2022computational,
  title={Computational imaging with dynamic metasurfaces: A recipe for simple and low-cost microwave imaging},
  author={Sleasman, Timothy and Imani, Mohammadreza F and Diebold, Aaron V and Boyarsky, Michael and Trofatter, Kenneth P and Smith, David R},
  journal={IEEE Antennas and Propagation Magazine},
  volume={64},
  number={4},
  pages={123--134},
  year={2022},
  publisher={IEEE}
}

@article{imani2020review,
  title={Review of metasurface antennas for computational microwave imaging},
  author={Imani, Mohammadreza F and Gollub, Jonah N and Yurduseven, Okan and Diebold, Aaron V and Boyarsky, Michael and Fromenteze, Thomas and Pulido-Mancera, Laura and Sleasman, Timothy and Smith, David R},
  journal={IEEE transactions on antennas and propagation},
  volume={68},
  number={3},
  pages={1860--1875},
  year={2020},
  publisher={IEEE}
}

@article{shlezinger2021dynamic,
  title={Dynamic metasurface antennas for 6G extreme massive MIMO communications},
  author={Shlezinger, Nir and Alexandropoulos, George C and Imani, Mohammadreza F and Eldar, Yonina C and Smith, David R},
  journal={IEEE Wireless Communications},
  volume={28},
  number={2},
  pages={106--113},
  year={2021},
  publisher={IEEE}
}

@article{yoo2018enhancing,
  title={Enhancing capacity of spatial multiplexing systems using reconfigurable cavity-backed metasurface antennas in clustered MIMO channels},
  author={Yoo, Insang and Imani, Mohammadreza F and Sleasman, Timothy and Pfister, Henry D and Smith, David R},
  journal={IEEE Transactions on Communications},
  volume={67},
  number={2},
  pages={1070--1084},
  year={2018},
  publisher={IEEE}
}

@article{sleasman2020implementation,
  title={Implementation and characterization of a two-dimensional printed circuit dynamic metasurface aperture for computational microwave imaging},
  author={Sleasman, Timothy A and Imani, Mohammadreza F and Diebold, Aaron V and Boyarsky, Michael and Trofatter, Kenneth P and Smith, David R},
  journal={IEEE Transactions on Antennas and Propagation},
  volume={69},
  number={4},
  pages={2151--2164},
  year={2020},
  publisher={IEEE}
}

@article{gros2020tuning,
  title={Tuning a regular cavity to wave chaos with metasurface-reconfigurable walls},
  author={Gros, Jean-Baptiste and del Hougne, Philipp and Lerosey, Geoffroy},
  journal={Physical Review A},
  volume={101},
  number={6},
  pages={061801},
  year={2020},
  publisher={APS}
}

@article{prod2024mutual,
  title={Mutual coupling in dynamic metasurface antennas: Foe, but also friend},
  author={Prod'homme, Hugo and del Hougne, Philipp},
  journal={arXiv preprint arXiv:2412.01002},
  year={2024}
}

@article{mazloum2024indoor,
  title={Indoor Channel Characterization using Transmitting and Reflecting RIS at mmWaves},
  author={Mazloum, Taghrid and Munoz, Fr{\'e}d{\'e}ric and Santamaria, Luca and Clemente, Antonio and Gros, Jean-Baptiste and Toubal, Ayoub and Nasser, Youssef and Odit, Mikhail and Lerosey, Geoffroy and D’errico, Raffaele},
  journal={IEEE Transactions on Antennas and Propagation},
  year={2024},
  publisher={IEEE}
}

@article{niu2019overview,
  title={An overview on backscatter communications},
  author={Niu, Jin-Ping and Li, Geoffrey Ye},
  journal={Journal of communications and information networks},
  volume={4},
  number={2},
  pages={1--14},
  year={2019},
  publisher={PTP}
}

@article{van2018ambient,
  title={Ambient backscatter communications: A contemporary survey},
  author={Van Huynh, Nguyen and Hoang, Dinh Thai and Lu, Xiao and Niyato, Dusit and Wang, Ping and Kim, Dong In},
  journal={IEEE Communications surveys \& tutorials},
  volume={20},
  number={4},
  pages={2889--2922},
  year={2018},
  publisher={IEEE}
}

@article{liang2022backscatter,
  title={Backscatter communication assisted by reconfigurable intelligent surfaces},
  author={Liang, Ying-Chang and Zhang, Qianqian and Wang, Jun and Long, Ruizhe and Zhou, Hu and Yang, Gang},
  journal={Proceedings of the IEEE},
  volume={110},
  number={9},
  pages={1339--1357},
  year={2022},
  publisher={IEEE}
}

@article{zhao2020metasurface,
  title={Metasurface-assisted massive backscatter wireless communication with commodity Wi-Fi signals},
  author={Zhao, Hanting and Shuang, Ya and Wei, Menglin and Cui, Tie Jun and Hougne, Philipp del and Li, Lianlin},
  journal={Nature communications},
  volume={11},
  number={1},
  pages={3926},
  year={2020},
  publisher={Nature Publishing Group UK London}
}

@inproceedings{zhang2017freerider,
  title={Freerider: Backscatter communication using commodity radios},
  author={Zhang, Pengyu and Josephson, Colleen and Bharadia, Dinesh and Katti, Sachin},
  booktitle={Proceedings of the 13th international conference on emerging networking experiments and technologies},
  pages={389--401},
  year={2017}
}

@article{popoff2014coherent,
  title={Coherent control of total transmission of light through disordered media},
  author={Popoff, S{\'e}bastien M and Goetschy, Arthur and Liew, SF and Stone, A Douglas and Cao, Hui},
  journal={Physical review letters},
  volume={112},
  number={13},
  pages={133903},
  year={2014},
  publisher={APS}
}

@article{del2016spatiotemporal,
  title={Spatiotemporal wave front shaping in a microwave cavity},
  author={Del Hougne, Philipp and Lemoult, Fabrice and Fink, Mathias and Lerosey, Geoffroy},
  journal={Physical review letters},
  volume={117},
  number={13},
  pages={134302},
  year={2016},
  publisher={APS}
}

@article{vellekoop2008phase,
  title={Phase control algorithms for focusing light through turbid media},
  author={Vellekoop, Ivo Micha and Mosk, AP},
  journal={Optics communications},
  volume={281},
  number={11},
  pages={3071--3080},
  year={2008},
  publisher={Elsevier}
}

@article{bouchet2021optimal,
  title={Optimal control of coherent light scattering for binary decision problems},
  author={Bouchet, Dorian and Rachbauer, Lukas M and Rotter, Stefan and Mosk, Allard P and Bossy, Emmanuel},
  journal={Physical Review Letters},
  volume={127},
  number={25},
  pages={253902},
  year={2021},
  publisher={APS}
}

\end{document}